\documentclass[twocolappendix,appendixfloats,]{emulateapj}

\usepackage[colorlinks = true, linkcolor = blue, urlcolor  = blue, citecolor = blue, anchorcolor = blue]{hyperref}
\usepackage{hyperref}
\usepackage{apjfonts}
\usepackage{rotating}
\usepackage{amsmath}
\usepackage{color}
\usepackage{graphicx}
\usepackage[dvipsnames]{xcolor}
\usepackage{CJK}

\newcommand{\te}{t_{\rm E}}
\newcommand{\thetae}{\theta_{\rm E}}

\newcommand{\pie}{\pi_{\rm E}}

\newcommand{\dl}{D_{\rm L}}

\definecolor{brown}{rgb}{0.59, 0.29, 0.0}
\definecolor{darkgreen}{rgb}{0.0, 0.42, 0.24}
\definecolor{darkblue}{rgb}{0.01, 0.31, 0.59}
\definecolor{darkpurple}{rgb}{1.25, 0.38, 2.05}

\def\eqalign#1{\null\,\vcenter{\openup\jot
        \ialign{\strut\hfil$\displaystyle{##}$&$
        \displaystyle{{}##}$\hfil \crcr#1\crcr}}\,}
\shortauthors{Han et al.}

\begin{document}

%\title{KMT-2019-BLG-1715LAB\lowercase{b}: microlensing giant planet in a binary stellar 
%system occurring on a binary Source }
\title{KMT-2019-BLG-1715: planetary microlensing event with three lens masses and two source stars}

\author{
% leading author -----------------------------
Cheongho~Han$^{1}$, 
Andrzej~Udalski$^{2}$, 
Doeon~Kim$^{1}$,
Youn~Kil~Jung$^{3}$, 
Chung-Uk~Lee$^{3}$, 
Ian~A.~Bond$^{4}$
\\
(Leading authors),\\
and \\
% KMTNet ---------------------------
Michael~D.~Albrow$^{0007}$, 
Sun-Ju~Chung$^{3,7}$,  
Andrew~Gould$^{8,9}$,
Kyu-Ha~Hwang$^{3}$,
Hyoun-Woo~Kim$^{3}$, 
Yoon-Hyun~Ryu$^{3}$, 
In-Gu~Shin$^{3}$, 
Yossi~Shvartzvald$^{10}$, 
Weicheng~Zang$^{11}$,
Jennifer~C.~Yee$^{4}$, 
Sang-Mok~Cha$^{3,12}$, 
Dong-Jin~Kim$^{3}$, 
Seung-Lee~Kim$^{3,7}$, 
Dong-Joo~Lee$^{3}$, 
Yongseok~Lee$^{3,12}$, 
Byeong-Gon~Park$^{3,7}$,
Richard~W.~Pogge$^{9}$,
Chun-Hwey Kim$^{13}$,
Woong-Tae Kim$^{14}$ \\
(The KMTNet Collaboration),\\
% OGLE -----------------------------
Przemek~Mr{\'o}z$^{2,15}$, 
Micha{\l}~K.~Szyma{\'n}ski$^{2}$,
Jan~Skowron$^{2}$,
Radek~Poleski$^{2}$, 
Igor~Soszy{\'n}ski$^{2}$,
Pawe{\l}~Pietrukowicz$^{2}$,
Szymon~Koz{\l}owski$^{2}$, 
Krzysztof~Ulaczyk$^{16}$,
Krzysztof~A.~Rybicki$^{2}$,
Patryk~Iwanek$^{2}$,
Marcin~Wrona$^{2}$, 
Mariusz~Gromadzki$^{2}$ \\
(The OGLE Collaboration) \\   
% ----
Fumio~Abe$^{17}$,                   
Richard~Barry$^{18}$,              
David~P.~Bennett$^{18,19}$,       
Aparna~Bhattacharya$^{18,19}$,    
Martin~Donachie$^{20}$,             
Hirosane~Fujii$^{17}$,              
Akihiko~Fukui$^{21,22}$,          
Yoshitaka~Itow$^{17}$,              
Yuki~Hirao$^{23}$,                  
Rintaro Kirikawa$^{23}$,
Iona~Kondo$^{23}$,                   
Man~Cheung~Alex~Li$^{20}$,           
Yutaka~Matsubara$^{16}$,             
Yasushi~Muraki$^{16}$,               
Shota~Miyazaki$^{23}$,               
Cl\'ement~Ranc$^{18}$,               
Nicholas~J.~Rattenbury$^{20}$,       
Yuki~Satoh$^{23}$,                   
Hikaru~Shoji$^{23}$,                 
Haruno~Suematsu$^{23}$,              
Takahiro~Sumi$^{23}$,                
Daisuke~Suzuki$^{23}$,               
Yuzuru Tanaka$^{23}$,
Paul~J.~Tristram$^{24}$,             
Takeharu~Yamakawa$^{16}$,            
Tsubasa~Yamawaki$^{23}$,             
Atsunori~Yonehara$^{25}$,\\                    
(The MOA Collaboration)\\
}

%\email{cheongho@astroph.chungbuk.ac.kr}

%=====================================
\affil{$^{1}$   Department of Physics, Chungbuk National University, Cheongju 28644, Republic of Korea,                                                      }
\affil{$^{2}$   Astronomical Observatory, University of Warsaw, Al.~Ujazdowskie 4, 00-478 Warszawa, Poland                                                   }
\affil{$^{3}$   Korea Astronomy and Space Science Institute, Daejon 34055, Republic of Korea                                                                 }
\affil{$^{4}$   Institute of Natural and Mathematical Sciences, Massey University, Auckland 0745, New Zealand                                                }
\affil{$^{5}$   Center for Astrophysics $|$ Harvard \& Smithsonian 60 Garden St., Cambridge, MA 02138, USA                                                   }
\affil{$^{6}$   University of Canterbury, Department of Physics and Astronomy, Private Bag 4800, Christchurch 8020, New Zealand                              }
\affil{$^{7}$   Korea University of Science and Technology, 217 Gajeong-ro, Yuseong-gu, Daejeon, 34113, Republic of Korea                                    }
\affil{$^{8}$   Max Planck Institute for Astronomy, K\"onigstuhl 17, D-69117 Heidelberg, Germany                                                             }
\affil{$^{9}$   Department of Astronomy, The Ohio State University, 140 W. 18th Ave., Columbus, OH 43210, USA                                                }
\affil{$^{10}$  Department of Particle Physics and Astrophysics, Weizmann Institute of Science, Rehovot 76100, Israel                                        }
\affil{$^{11}$  Department of Astronomy and Tsinghua Centre for Astrophysics, Tsinghua University, Beijing 100084, China                                     }
\affil{$^{12}$  School of Space Research, Kyung Hee University, Yongin, Kyeonggi 17104, Republic of Korea                                                    }
\affil{$^{13}$  Department of Astronomy \& Space Science, Chungbuk National University, Cheongju 28644, Republic of Korea                                    }
\affil{$^{14}$  Department of Physics \& Astronomy, Seoul National University, Seoul 08826, Republic of Korea                                                }
\affil{$^{15}$  Division of Physics, Mathematics, and Astronomy, California Institute of Technology, Pasadena, CA 91125, USA                                 }
\affil{$^{16}$  Department of Physics, University of Warwick, Gibbet Hill Road, Coventry, CV4 7AL, UK                                                        }
\affil{$^{17}$  Institute for Space-Earth Environmental Research, Nagoya University, Nagoya 464-8601, Japan                                                  }
\affil{$^{18}$  Code 667, NASA Goddard Space Flight Center, Greenbelt, MD 20771, USA                                                                         }
\affil{$^{19}$  Department of Astronomy, University of Maryland, College Park, MD 20742, USA                                                                 }
\affil{$^{20}$  Department of Physics, University of Auckland, Private Bag 92019, Auckland, New Zealand                                                      }
\affil{$^{21}$  Instituto de Astrof\'isica de Canarias, V\'ia L\'actea s/n, E-38205 La Laguna, Tenerife, Spain                                               }
\affil{$^{22}$  Department of Earth and Planetary Science, Graduate School of Science, The University of Tokyo, 7-3-1 Hongo, Bunkyo-ku, Tokyo 113-0033, Japan}
\affil{$^{23}$  Department of Earth and Space Science, Graduate School of Science, Osaka University, Toyonaka, Osaka 560-0043, Japan                         }
\affil{$^{24}$  University of Canterbury Mt.\ John Observatory, P.O. Box 56, Lake Tekapo 8770, New Zealand                                                   }
\affil{$^{25}$  Department of Physics, Faculty of Science, Kyoto Sangyo University, 603-8555 Kyoto, Japan                                                    }
%\altaffiltext{100}{KMTNet Collaboration.}
%\altaffiltext{101}{OGLE Collaboration.}
%\altaffiltext{102}{MOA Collaboration.}

\begin{abstract}
We investigate the gravitational microlensing event KMT-2019-BLG-171, of which light curve shows
two short-term anomalies from a caustic-crossing binary-lensing light curve: one with a large 
deviation and the other with a small deviation.  We identify five pairs of solutions, in which the 
anomalies are explained by adding an extra lens or source component in addition to the base binary-lens 
model.  We resolve the degeneracies by applying a method, in which the measured flux ratio between the 
first and second source stars is compared with the flux ratio deduced from the ratio of the source radii.  
Applying this method leaves a single pair of viable solutions, in both of which the major anomaly is 
generated by a planetary-mass third body of the lens, and the minor anomaly is generated by a faint 
second source.  A Bayesian analysis indicates that the lens comprises three masses: a planet-mass 
object with $\sim 2.6~M_{\rm J}$ and binary stars of K and M dwarfs lying in the galactic disk.  
We point out the possibility that the lens is the blend, and this can be verified by conducting 
high-resolution followup imaging for the resolution of the lens from the source.

\end{abstract}

\keywords{Gravitational microlensing (672); Gravitational microlensing exoplanet detection (2147)}
% Brown dwarfs (185)
% Binary stars (154)
% Free floating planets (549)

\section{Introduction}\label{sec:one}

During the first-generation experiment \citep{Udalski1994a, Alcock1997}, microlensing observations 
were carried out with about one day cadence.  By employing wide-field cameras mounted on multiple 
telescopes, the cadence of microlensing observations has been dramatically shortened, and now it 
reaches 15 minutes for the fields of the highest stellar concentration.  With the shortened cadence, 
the event detection rate has greatly increased from several dozens/yr during the first-generation 
experiments to the current rate of more than 3000 events/yr.

Light curves of lensing events often deviate from that of a single lens and a single source (1L1S) 
event, which produces a smooth and symmetric light curve.  The most common cause of the deviation 
is the binarity of the lens: 2L1S events \citep{Mao1991}.  Interpreting the light curves of 2L1S 
events was a challenging task at the time when such events were first detected, e.g., OGLE No.~7 
\citep{Udalski1994b}, because the methodology for the analysis of these anomalous events had not 
yet been developed.  With the development of various methodologies, e.g., ray-shooting technique
\citep{Bond2002, Dong2009, Bennett2010} and contour integration algorithm \citep{Gould1997, 
Bozza2018} developed for finite magnification computations, followed by the theoretical understanding 
of the binary lensing physics, 2L1S events are routinely  detected and analyzed as they progress 
\citep{Ryu2010, Bozza2012}.

With the increasing number of densely covered events, one occasionally confronts events having 
light curves that exhibit extra deviations from those of 2L1S events.  One important cause of 
such extra deviations is the existence of an additional lens or source component.  There exist 
12 confirmed events, for which at least four objects, including the lens and source components, 
are needed for the interpretations of the light curves.  See Table~1 of \citet{Han2021}.  Among 
these events, 9 events were produced by triple lens systems, 3L1S events.  For two major 
reasons, modeling a 3L1S event is difficult.  The first is the complexity of the lensing behavior 
\citep{Danek2015, Danek2019} and the resulting difficulties in analyzing deviations. Another 
important obstacle in analyzing these events is the degeneracy problem, which makes it difficult 
to find a correct solution among those resulting in similar light curves despite dramatically different 
interpretations of the lens system.  Identifying various types of degeneracies and investigating 
their causes are important not only to correctly interpret the lens system but also to investigate 
similar degeneracies in following analyses.

We investigate the microlensing event KMT-2019-BLG-1715 and present  the analysis result.  The 
event displays a light curve with a complex pattern, in which there exist two short-lasting 
anomalies deviating from a caustic-crossing 2L1S light curve.  We model the observed light 
curve under various interpretations and present the results.

The organization of the paper for the presentation of the analysis is as follows.  We address 
the acquisition and reduction of data in Sect.~\ref{sec:two}.  We depict the pattern of the 
anomalies that deviate from a 2L1S lensing light curve in Sect.~\ref{sec:three}.  We present 
various models describing the anomalies in Sect.~\ref{sec:four}.  We inspect the origins of the 
degeneracies among the identified solutions in Section~\ref{sec:five}.  We characterize the source, 
and measure the Einstein radius in Sect.~\ref{sec:six}.  We probe a method that can resolve the 
degeneracies in the lensing models in Sect.~\ref{sec:seven}.  In Sect.~\ref{sec:eight}, we describe 
a Bayesian analysis conducted to characterize the lens, and present the physical quantifies.  We 
summarize the analysis and make a conclusion in Sect.~\ref{sec:nine}.

% Table 1 ------------------------------------------------
\begin{deluxetable}{lcccc}
\tablecaption{Data and error readjustment factors\label{table:one}}
\tablewidth{240pt}
%\tabletypesize{\small}
\tablehead{
\multicolumn{1}{c}{Data set}                           &
\multicolumn{1}{c}{$k$}                                &
\multicolumn{1}{c}{$\sigma_{\rm min}$ (mag)}           &
\multicolumn{1}{c}{$N_{\rm data}$}                     
}
\startdata                                 
OGLE           & 1.225    &  0.030    &    450    \\
KMTA (BLG03)   & 1.404    &  0.020    &   1307    \\
KMTA (BLG43)   & 0.812    &  0.050    &   1195    \\
KMTA (BLG03)   & 1.517    &  0.010    &   1109    \\
KMTA (BLG43)   & 1.217    &  0.020    &   1424    \\
KMTC (BLG03)   & 1.382    &  0.010    &    981    \\
KMTS (BLG43)   & 1.187    &  0.020    &   1314    
\enddata                            
%\tablecomments{ ${\rm HJD}^\prime = {\rm HJD}- 2450000$.  
\smallskip
%}
\end{deluxetable}
% --------------------------------------------------------

% Figure 1 ------------------------------------------------------
\begin{figure}
\includegraphics[width=\columnwidth]{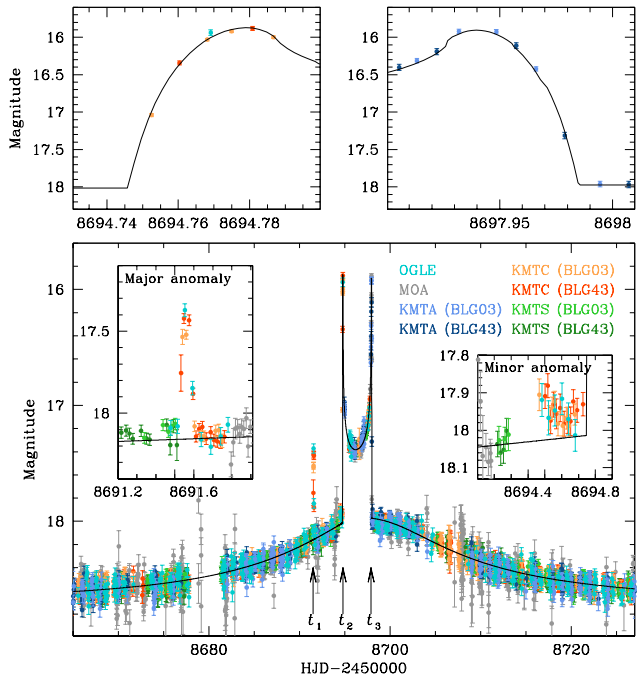}
\caption{
Lensing light curve of KMT-2019-BLG-1715.  The model obtained from a 2L1S interpretation 
(close model) is drawn over the data points.  The two top panels show the zoom-in views around 
the times of the source star's entrance into and exit from the caustic according to the 2L1S model.  
The two insets in the lower panel show the zoomed-in views around the two anomalies from the 2L1S 
model occurred at ${\rm HJD}^\prime\sim 8691.5$ (major anomaly) and $\sim 8694.5$ (minor anomaly).  
\smallskip
}
\label{fig:one}
\end{figure}
% --------------------------------------------------------------

\section{Observations and Data}\label{sec:two}

The event KMT-2019-BLG-1715 was found from the observations of a source star lying 
at $({\rm RA}, {\rm DEC})_{\rm J2000} = (18:01:29.21, -28:46:37.7)$, that correspond to 
$(l, b) = (1^\circ\hskip-2pt.898, -2^\circ\hskip-2pt.914)$.  
The apparent source magnitude at the baseline was $I_{\rm base}=18.65$.

The magnification of the source flux was discovered from 
the Korea Microlensing Telescope Network \citep[KMTNet:][]{Kim2016} survey on 2019 July 20 
(${\rm HJD}^\prime \equiv {\rm HJD}-2450000\sim 8684.7$), at which the source became brighter 
by about $0.25$ magnitude from the baseline.  The KMTNet survey utilizes three telescopes that 
lie around the world in Australia (KMTA), Chile (KMTC), and South Africa (KMTS).  The aperture 
of each telescope is 1.6 meter, and the field of view of the camera mounted on the telescope is 
$4~{\rm deg}^2$.  The source lies in the BLG03 and BLG43 fields, which overlap with a small 
offset, and thus the data are composed of six sets, with two sets from each telescope.

The event was independently found by two other lensing surveys of the Optical Gravitational 
Lensing Experiment \citep[OGLE:][]{Udalski2015} and the Microlensing Observations in Astrophysics 
\citep[MOA:][]{Bond2001} on 2019 July 28 (${\rm HJD}^\prime \sim 8692.7$) and August 1 
(${\rm HJD}^\prime \sim 8696.7$), respectively.  The OGLE and MOA surveys designated the event 
as OGLE-2019-BLG-1190 and MOA-2019-BLG-352, respectively.  The OGLE survey uses the 1.3~m Warsaw 
Telescope located at the Las Campanas Observatory in Chile, and the MOA survey utilizes the 1.8~m 
MOA-II Telescope located at Mt.~John Observatory in New Zealand.

% Figure 2 ------------------------------------------------------
\begin{figure}
\includegraphics[width=\columnwidth]{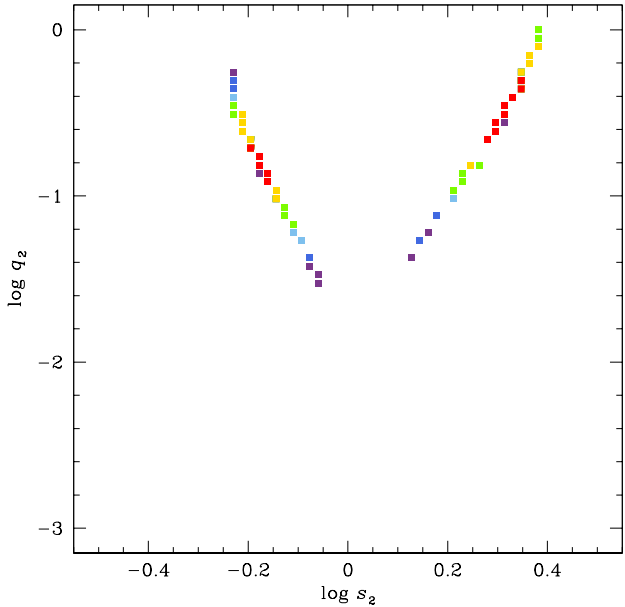}
\caption{
Distribution of $\Delta\chi^2$ on the $\log s$--$\log q$ plane.  Red, yellow, green, cyan, 
blue, and purple colors are used to designate points with $<1n\sigma$, $<2n\sigma$, $<3n\sigma$, 
$<4n\sigma$, $<5n\sigma$, and $<6n\sigma$, where $n=10$.
\smallskip
}
\label{fig:two}
\end{figure}
% --------------------------------------------------------------

The data were processed utilizing the photometry pipelines of the individual groups: 
\citet{Albrow2017}, \citet{Udalski2003}, and \citet{Bond2001} for the KMTNet, OGLE, and MOA 
surveys, respectively.  We note that these pipelines commonly utilize the difference imaging 
algorithm that was developed to optimize photometry in very crowded fields \citep{Alard1998}.  
Although the MOA data cover  the caustic crossing at ${\rm HJD}^\prime\sim 8697.9$, 
which is also densely covered by the KMTA data sets, they do not cover the anomalies of our major 
interest to be discussed below.  Furthermore, the photometric uncertainties of the data are 
substantially larger than those of the other data sets, and thus the MOA data are not used for 
analysis to minimize their effects on modeling.  The error bars of the data used in the analysis 
were rescaled from those estimated from the automatized pipelines, $\sigma_0$, by 
$\sigma=k(\sigma_{\rm min}^2+\sigma_0^2)^{1/2}$, where $k$ denotes a rescaling factor to normalize 
$\chi^2/{\rm dof}$, where dof represents the degree of freedom, to unity, and $\sigma_{\rm min}$ 
is the scatter of data \citep{Yee2012}.  In Table~\ref{table:one}, we list $(k, \sigma_{\rm min}, 
N_{\rm data})$ for the six data sets.  Here $N_{\rm data}$ denotes the number of data 
in each data set.

% Table 2 ------------------------------------------------
\begin{deluxetable}{lccc}
\tablecaption{Lensing parameters of 2L1S model\label{table:two}}
\tablewidth{240pt}
%\tabletypesize{\small}
\tablehead{
\multicolumn{1}{c}{Parameter}           &
\multicolumn{1}{c}{Close}               &
\multicolumn{1}{c}{Wide}                
}
\startdata                                 
$\chi^2$                    &   9329.3                  &   9313.7                  \\ 
$t_0$ (${\rm HJD}^\prime$)  &   $8697.039 \pm 0.022$    &   $8696.971 \pm 0.016$    \\
$u_0$                       &   $0.123 \pm 0.002   $    &   $0.091 \pm 0.001   $    \\ 
$\te$ (days)                &   $30.94\pm 0.31     $    &   $39.04 \pm 0.35    $    \\
$s$                         &   $0.756 \pm 0.006   $    &   $2.028 \pm 0.002   $    \\
$q$                         &   $0.141 \pm 0.002   $    &   $0.310 \pm 0.006   $    \\
$\alpha$ (rad)              &   $4.945 \pm 0.005   $    &   $4.913 \pm 0.005   $    \\ 
$\rho$ ($10^{-3}$)          &   $0.65 \pm 0.01     $    &   $0.57 \pm 0.01     $    
\enddata                            
%\tablecomments{ ${\rm HJD}^\prime = {\rm HJD}- 2450000$.  \smallskip }
\end{deluxetable}
% --------------------------------------------------------

% Figure 3 ------------------------------------------------------
\begin{figure}
\includegraphics[width=\columnwidth]{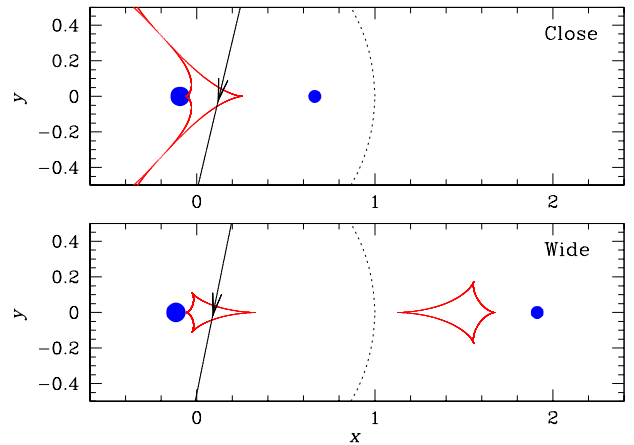}
\caption{
Lensing configurations of the two 2L1S models. The blue dots indicate the lens positions, 
the red figures represent the caustics, and the arrowed line represents the source 
motion.  The Einstein ring is marked by a dotted circle. 
\bigskip
}
\label{fig:three}
\end{figure}
% --------------------------------------------------------------

\section{2L1S interpretation}\label{sec:three}

The observed light curve of KMT-2019-BLG-1715 is shown in Figure~\ref{fig:one}.  At first 
glance, the two pronounced caustic-crossing spikes, at ${\rm HJD}^\prime\sim 8694.7$ ($t_2$) 
and $\sim 8697.9$ ($t_3$), suggests that the event is a usual caustic-crossing 2L1S event.  
Fitting the light curve with a 2L1S model yields a pair of solutions with $(s, q)\sim (0.76, 0.14)$ 
(``close'' model) and $(s, q)\sim (2.03, 0.31)$ (``wide'' model), where $s$ and $q$ denote the 
projected separation (scaled to the Einstein radius $\thetae$) and the mass ratio between the binary 
lens components, with masses $M_1$ and $M_2$, respectively.  We carry out the 2L1S modeling in two 
steps, in which the binary parameters $(s,q)$ are searched for via a dense grid approach in the first 
step, and locals appearing in the $\Delta\chi^2$ map are refined in the second step.  The ranges of 
the grid parameters are $-1.0 \leq \log s \leq 1.0$ and $-3.0 \leq \log q \leq 0.0$ with 60 divisions.  
Figure~\ref{fig:two} shows the $\Delta\chi^2$ distribution obtained from the grid search and the 
locations of the two locals, i.e., close and wide solutions.

% Figure 4 ------------------------------------------------------
\begin{figure}
\includegraphics[width=\columnwidth]{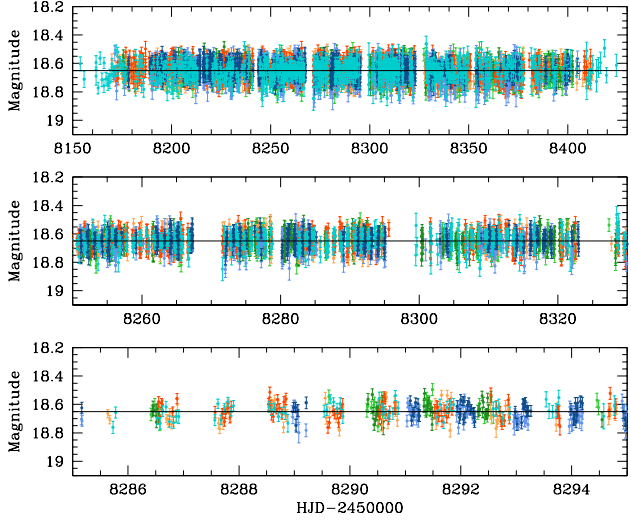}
\caption{
Light curves at the baseline during three different time spans: 
280~days (top panel), 80~days (middle panel), and 10~days (bottom panel).
The colors of data points are same as those in Fig.~\ref{fig:one}.
\smallskip
}
\label{fig:four}
\end{figure}
% --------------------------------------------------------------

We list the full lensing parameters of the two 2L1S solutions in Table~\ref{table:two}.  In 
the table, the parameters $(t_0, u_0, \te)$ denote the closest source approach time to a
lens reference position, the reference-source separation at $t_0$, and the timescale of the 
event, respectively.  As a lens reference, we choose the center of mass for a close binary 
($s<1.0$), and the effective lens position, defined by \citet{Stefano1996} and \citet{An2002}, 
for a wide binary ($s>1.0$).  Because the light curve exhibits pronounced caustic-crossing spikes, 
finite-source effects (hereafter ``finite effects'') are considered in modeling by including 
$\rho=\theta_*/\thetae$ (normalized ``source radius'') as an additional parameter.  Here $\theta_*$ 
denote the angular source radius.  Hereafter, we denote $\theta_*$ and $\thetae$ without the notation 
of ``angular''.  In the computation of finite magnifications, we take limb-darkening effects into 
consideration assuming that the brightness profile on the source surface varies as 
$\Sigma_\lambda \propto 1- (1-3\cos\phi/2)\Gamma_\lambda$.  Here $\Gamma_\lambda$ denotes the 
coefficient of limb darkening, and $\phi$ denotes the angle between the two lines extending from 
the source center, one toward an observer and the other toward the source surface.  Based on the 
spectral type, to be discussed in Section~\ref{sec:six}, we adopt the $V$- and $I$-band coefficients 
of $(\Gamma_V, \Gamma_I)=(0.62, 0.45)$.  The two solutions (close and wide) are degenerate, and the 
degeneracy is originated from the close--wide degeneracy known by \citet{Dominik1999}.  The lensing 
configurations, showing the trajectory of the source relative to the lens components and caustic, 
are provided in Figure~\ref{fig:three}.  The model curve of the 2L1S solution (for the close model) 
is shown in Figure~\ref{fig:one}.

% Figure 5 ------------------------------------------------------
\begin{figure}
%\centering
\includegraphics[width=\columnwidth]{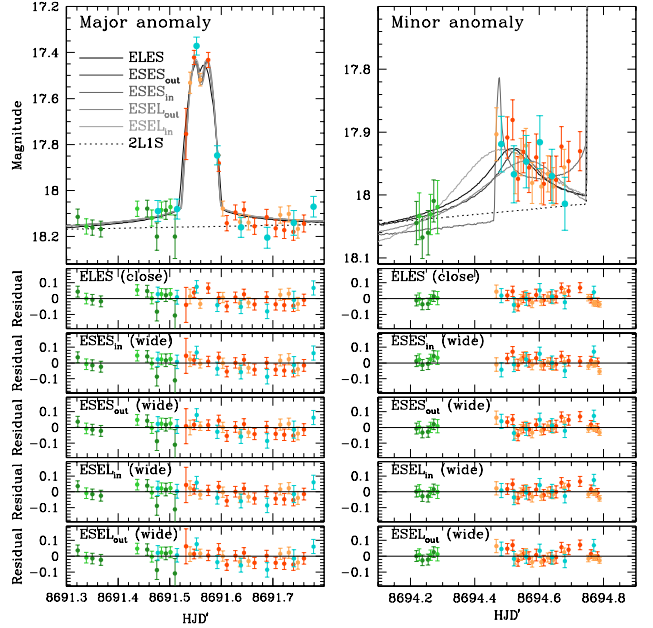}
\caption{
Zoomed-in views of the light curve around the major (at around $t_1$) and minor (at around $t_2$) 
anomalies.  For each epoch, the lower five panels show the residuals from the identified five sets of 
degenerate solutions: ELES (close), ESES$_{\rm in}$ (wide), ESEL$_{\rm out}$ (wide), ESEL$_{\rm in}$ 
(wide), and ESEL$_{\rm out}$ (wide) solutions. For each set, there are two solutions with 
$s_2<1.0$ (close) and $s_2>1.0$ (wide), and the presented residuals are for the solution yielding 
a better fit among the close--wide solutions.  Drawn over the data points are models of five 
degenerate solutions.  The degeneracies among the solutions are severe, resulting in similar model 
curves.  The only significant difference is the extra caustic entrance at $t_2$ corresponding to 
the ESES$_{\rm in}$ model. 
\smallskip
}
\label{fig:five}
\end{figure}
% --------------------------------------------------------------

Although the 2L1S model approximately describes the overall features of the observed light 
curve, a close look reveals that the data exhibit two short-lasting anomalies from the 
base 2L1S model.  The first anomaly appears at ${\rm HJD}^\prime\sim 8691.5$ ($t_1$) with 
$\Delta I\sim 0.7$~mag deviation from the 2L1S model.  The second anomaly appears just before the 
second caustic crossing at $\sim 8694.5$ ($t_2$) with $\Delta I\sim 0.1$~mag deviation.  We refer 
to the former and latter anomalies as the ``major anomaly'' and ``minor anomaly'', respectively.  
The two insets in the lower panel of Figure~\ref{fig:one} show the zoom-in views of the anomalies.
The major anomaly exhibits both rising and falling parts during its very short duration of $\sim 3$ hours.  
The duration of the minor anomaly is short as well, but precisely estimating the duration is difficult 
because the coverage of the anomaly is incomplete.

We check the possibility that the extra anomalies arise due to photometric artifacts or stellar 
variability.  First, the possibility of the photometric artifact, such as the change in transparency 
or the passage of a Solar System object across the source, is very unlikely because both anomalies 
are covered by multiple remotely separated telescopes and the data during the anomalies show 
consistent patterns of variations.  Second, the possibility of the source variability is also 
unlikely because the stellar type of the source is a main sequence, for which the chance of light 
variation is very small.  We check the source variability by inspecting the light curve at the baseline.  
Figure~\ref{fig:four} shows the baseline light curve during three different time spans: 280~days 
(top panel), 80~days (middle panel), and 10~days (bottom panel).  It is found that the light curve 
does not show any variation in all inspected time ranges.

% Table 3 ------------------------------------------------
\begin{deluxetable*}{lccccc}
\tablecaption{Lensing parameters of ESES models\label{table:three}}
\tablewidth{480pt}
%\tabletypesize{\small}
\tablehead{
\multicolumn{1}{c}{Parameter}          &
\multicolumn{2}{c}{ESES$_{\rm in}$}    &
\multicolumn{2}{c}{ESES$_{\rm out}$}   \\
\multicolumn{1}{c}{}                   &
\multicolumn{1}{c}{Close}              &
\multicolumn{1}{c}{Wide}               &
\multicolumn{1}{c}{Close}              &
\multicolumn{1}{c}{Wide}     
}
\startdata               
$\chi^2$                            &   7737.0                 &   7696.1                 &   7771.2                 &   7701.1                  \\ 
$t_{0,1}$  (${\rm HJD}^\prime$)     &   $8697.047 \pm  0.019$  &   $8697.153 \pm  0.017$  &  $8697.046 \pm  0.015$   &   $8697.113 \pm  0.016$   \\
$u_{0,1}$                           &   $0.103 \pm  0.002   $  &   $0.107 \pm  0.001   $  &  $0.102 \pm  0.001   $   &   $0.096 \pm  0.001   $   \\
$t_{0,2}$  (${\rm HJD}^\prime$)     &   $8693.518 \pm  0.035$  &   $8694.716 \pm 0.059 $  &  $8693.525 \pm  0.029$   &   $8694.400 \pm  0.052$   \\
$u_{0,2}$                           &   $0.205 \pm  0.002   $  &   $0.368 \pm  0.002   $  &  $0.204 \pm  0.002   $   &   $0.304 \pm  0.002   $   \\
$t_{0,3}$  (${\rm HJD}^\prime$)     &   $8696.383 \pm  0.037$  &   $8697.270 \pm  0.058$  &  $8696.669 \pm  0.047$   &   $8697.865 \pm  0.105$   \\
$u_{0,3}$                           &   $0.188 \pm  0.003   $  &   $0.309 \pm  0.003   $  &  $0.222 \pm  0.003   $   &   $0.357 \pm  0.007   $   \\
$\te$ (days)                        &   $32.58 \pm  0.28    $  &   $35.43 \pm  0.17    $  &  $32.66 \pm  0.05    $   &   $37.31 \pm  0.25    $   \\
$s$                                 &   $0.696 \pm  0.004   $  &   $1.988 \pm  0.001   $  &  $0.693 \pm  0.003   $   &   $2.078 \pm  0.003   $   \\
$q$                                 &   $0.177 \pm  0.003   $  &   $0.349 \pm  0.004   $  &  $0.178 \pm  0.002   $   &   $0.385 \pm  0.007   $   \\
$\alpha$ (rad)                      &   $4.997 \pm  0.005   $  &   $4.950 \pm  0.004   $  &  $4.999 \pm  0.004   $   &   $4.957 \pm  0.005   $   \\
$\rho_1$ ($10^{-3}$)                &   $0.59 \pm  0.01     $  &   $0.63 \pm  0.01     $  &  $0.60 \pm  0.01     $   &   $0.58 \pm  0.01     $   \\
$\rho_2$ ($10^{-3}$)                &   $0.47 \pm  0.04     $  &   $0.45 \pm  0.04     $  &  $0.42 \pm  0.04     $   &   $0.43 \pm  0.04     $   \\
$\rho_3$ ($10^{-3}$)                &    --                    &    --                    &   --                     &    --                     \\
$q_{F,2}$                           &   $0.051 \pm  0.002   $  &   $0.060 \pm  0.002   $  &  $0.050 \pm  0.001   $   &   $0.059 \pm  0.002   $   \\
$q_{F,3}$                           &   $0.020 \pm  0.003   $  &   $0.021 \pm  0.003   $  &  $0.029 \pm  0.004   $   &   $0.040 \pm  0.006   $   
\enddata                            
%\tablecomments{ ${\rm HJD}^\prime \equiv {\rm HJD}-2450000$.  }
\end{deluxetable*}
% --------------------------------------------------------

\section{Interpreting the anomalies}\label{sec:four}

We investigate lensing models that can explain the observed anomalies from the 2L1S models.  From 
this investigation, we find that adding a fourth body (either a source or a lens) can explain one 
of the two anomalies, but not both simultaneously.  Solutions explaining both anomalies can only 
be found from combinations of an extra lens (EL) and an extra source (ES), and we find five sets 
of such solutions resulting from various types of degeneracy.  For each set, there exist two solutions 
with $s_2>1.0$ and $s_2<1.0$ resulting from the close--wide binary degeneracy, and thus there are ten 
solutions in total.  The details of the individual models are discussed in the following subsections.

For the designation of the individual solutions to be discussed, we use the notations 
``ELES'' (extra lens and extra source), ``ESES'' (extra source and extra source), and 
``ESEL'' (extra source and extra lens). In these notations, the {\it first} ``L'' (or 
``S'') indicates that an extra lens (an extra source) is included in the model to 
explain the {\it major} anomaly, while the {\it second} ``L'' (or ``S'') denotes that 
an extra lens (or source) is included to describe the {\it minor} anomaly. For example, 
``ELES'' denotes a model, in which the major and minor anomalies are explained by 
adding an extra lens and an extra source to the base 2L1S solution, respectively, 
and thus the model consists of three lens components and two source stars.

% Figure 6 ------------------------------------------------------
\begin{figure}
\includegraphics[width=\columnwidth]{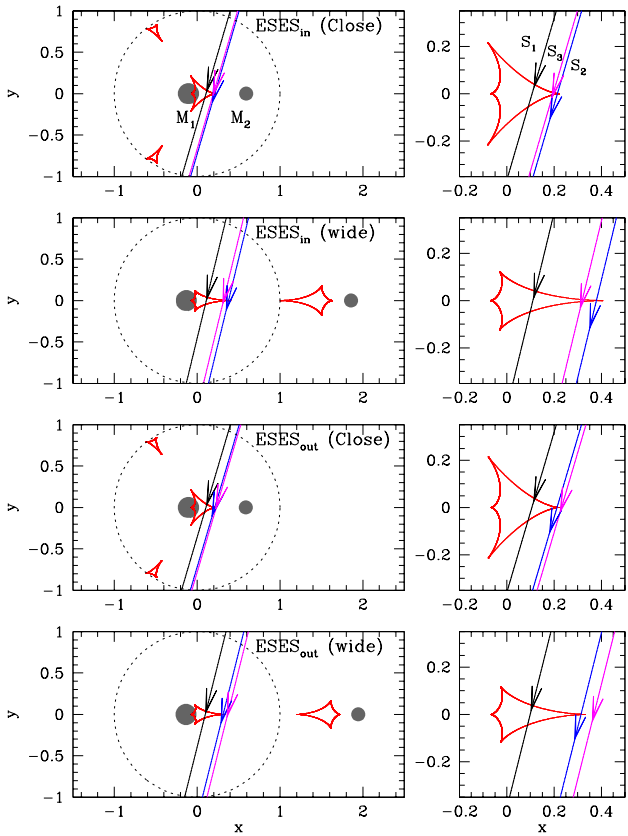}
\caption{
Lensing configurations for the four ESES models: 
ESES$_{\rm in}$ (close), 
ESES$_{\rm in}$ (wide), 
ESES$_{\rm out}$ (close), and
ESES$_{\rm out}$ (wide). 
For each solution, the panel on the left shows the wide view including the lens positions, and 
the panel on the right shows the central region.  The positions of the lens components are marked 
by two filled dots ($M_1$ and $M_2$).  The three lines with arrows denote the motion of the primary 
($S_1$, black), second ($S_1$, blue), and third ($S_1$, magenta) source stars.  
\bigskip
}
\label{fig:six}
\end{figure}
% --------------------------------------------------------------

\subsection{ESES solutions}\label{sec:four-one}

The ESES solutions explain both the major and minor anomalies with the addition of two extra 
source stars to the base 2L1S model.  In the modeling considering additional source stars, the 
initial lensing parameters related to the first source are adopted from those of the 2L1S model, 
and the initial parameters related to the other source stars are assigned considering the locations 
and magnitudes of the anomalies.  According to these solutions, then, there are two lens masses 
($M_1$ and $M_2$) and three source stars ($S_1$, $S_2$, and $S_3$): 2L3S model.  Besides the 
close-wide degeneracy, the ESES interpretation additionally suffers from a degeneracy arising 
due to the ambiguity in the trajectory of $S_3$, resulting in four degenerate solutions in total.  
The two sets of the degenerate solutions resulting from the latter type degeneracy,  ESES$_{\rm in}$ 
and ESES$_{\rm out}$, commonly explain the origin of the major anomaly as the second source ($S_2$) 
star's caustic crossing over the tip of the binary-induced caustic.  According to the ESES$_{\rm in}$ 
and ESES$_{\rm out}$ solutions, the tertiary source ($S_3$) star, that is introduced to explain the 
minor anomaly, passes the {\it inner} and {\it outer} region of the caustic, respectively.  The 
degeneracy between the two solutions results from the incomplete coverage of the minor anomaly.

% Table 4 ------------------------------------------------
\begin{deluxetable*}{lccccc}
\tablecaption{Lensing parameters of ESEL models\label{table:four}}
\tablewidth{480pt}
%\tabletypesize{\small}
\tablehead{
\multicolumn{1}{c}{Parameter}          &
\multicolumn{2}{c}{ESEL$_{\rm in}$}     &
\multicolumn{2}{c}{ESEL$_{\rm out}$}   \\
\multicolumn{1}{c}{}    &
\multicolumn{1}{c}{Close}    &
\multicolumn{1}{c}{Wide}    &
\multicolumn{1}{c}{Close}    &
\multicolumn{1}{c}{Wide}    
}
\startdata                           
$\chi^2$                           &  7713.6                 &   7668.5                 &   7741.3                  &    7667.9                 \\ 
$t_{0,1}$ (${\rm HJD}^\prime$)     &  $8697.094 \pm  0.015$  &   $8697.071 \pm 0.016$   &   $8697.141 \pm 0.014$    &    $8697.023 \pm 0.015$   \\
$u_{0,1}$                          &  $0.109 \pm  0.002   $  &   $0.092 \pm 0.001   $   &   $0.121 \pm 0.001   $    &    $0.078 \pm 0.002   $   \\
$t_{0,2}$ (${\rm HJD}^\prime$)     &  $8693.547 \pm  0.034$  &   $8694.239 \pm 0.057$   &   $8693.599 \pm 0.033$    &    $8693.982 \pm 0.066$   \\
$u_{0,2}$                          &  $0.216 \pm  0.003   $  &   $0.292 \pm 0.004   $   &   $0.232 \pm 0.002   $    &    $0.229 \pm 0.010   $   \\
$\te$ (days)                       &  $32.05 \pm  0.39    $  &   $38.35 \pm 0.34    $   &   $30.57 \pm 0.28    $    &    $42.92 \pm 1.03    $   \\
$s_2$                              &  $0.716 \pm  0.004   $  &   $2.089 \pm 0.009   $   &   $0.739 \pm 0.003   $    &    $2.227 \pm 0.026   $   \\
$q_2$                              &  $0.164 \pm  0.002   $  &   $0.372 \pm 0.007   $   &   $0.162 \pm 0.002   $    &    $0.398 \pm 0.012   $   \\
$\alpha$ (rad)                     &  $4.991 \pm  0.005   $  &   $4.947 \pm 0.005   $   &   $4.992 \pm 0.005   $    &    $4.954 \pm 0.004   $   \\
$s_3$                              &  $1.178 \pm  0.006   $  &   $0.952 \pm 0.004   $   &   $1.098 \pm 0.003   $    &    $0.891 \pm 0.005   $   \\
$q_3$ ($10^{-3}$)                  &  $0.26 \pm  0.05     $  &   $0.06 \pm 0.01     $   &   $0.05 \pm 0.01     $    &    $0.03 \pm 0.01     $   \\
$\psi$ (rad)                       &  $0.734 \pm  0.004   $  &   $0.902 \pm 0.003   $   &   $0.711 \pm 0.003   $    &    $0.929 \pm 0.005   $   \\
$\rho_1$ ($10^{-3}$)               &  $0.62 \pm  0.01     $  &   $0.57 \pm 0.01     $   &   $0.65 \pm 0.01     $    &    $0.50 \pm 0.02     $   \\
$\rho_2$ ($10^{-3}$)               &  $0.52 \pm  0.04     $  &   $0.43 \pm 0.04     $   &   $0.48 \pm 0.04     $    &    $0.34 \pm 0.04     $   \\
$q_F$                              &  $0.052 \pm  0.002   $  &   $0.059 \pm 0.002   $   &   $0.049 \pm 0.002   $    &    $0.057 \pm 0.002   $   
\enddata                                                       
%\tablecomments{ ${\rm HJD}^\prime \equiv {\rm HJD}-2450000$.  \bigskip}
\end{deluxetable*}
% --------------------------------------------------------

The lensing configurations of the four ESES solutions are presented in Figure~\ref{fig:six}, 
in which the trajectories of the second and third source stars are marked by $S_2$, and $S_3$, 
respectively.  The model curves and residuals from the models in the regions around the anomalies 
are shown in Figure~\ref{fig:five}.  We note that the presented models are the wide solutions, 
which yield better fits than the corresponding close solutions.  The minor anomaly according to 
the ESES$_{\rm in}$ solution is characterized by a U-shape trough pattern because $S_3$ passes the 
inner region of the caustic.   The lensing parameters of the ESES$_{\rm in}$ and ESES$_{\rm out}$, 
including both the close and wide solutions, are listed in Table~\ref{table:three}.  The flux ratios 
of the secondary and tertiary source stars to the primary source are 
$q_{F,2}=F_{S,2}/F_{S,1} \sim 0.05$--0.06 and $q_{F,3}=F_{S,3}/F_{S,1}\sim 0.02$--0.05, respectively.  
Here $F_{S,i}$ and $(t_{0,i}, u_{0,i}, \rho_i)$ indicate the flux and the lensing parameters related 
to the $i$th source star, respectively.  It is found that $\rho_3$ cannot be firmly determined because 
the coverage of the minor anomaly is incomplete.

% Figure 7 ------------------------------------------------------
\begin{figure}
\includegraphics[width=\columnwidth]{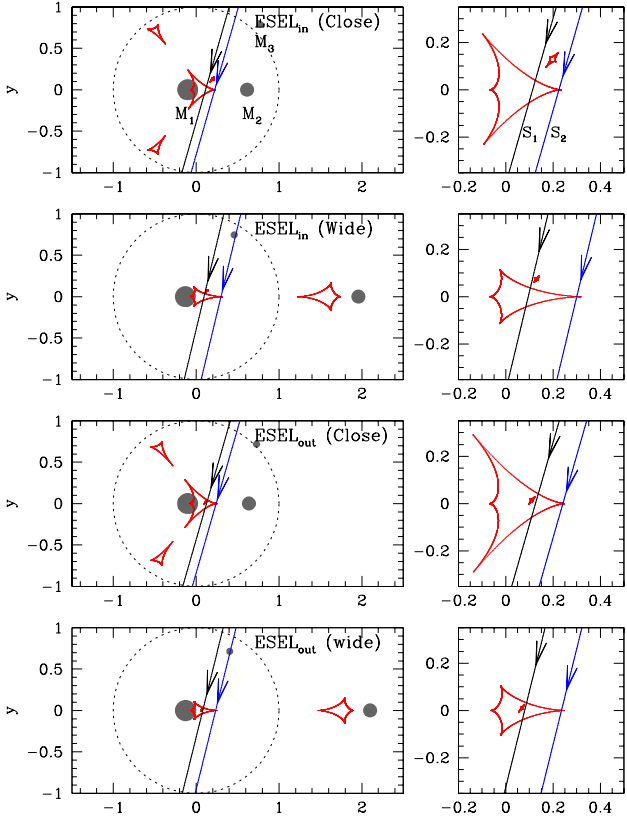}
\caption{
Lensing configurations of the four ESEL solutions: 
ESEL$_{\rm in}$ (close), 
ESEL$_{\rm in}$ (wide), 
ESEL$_{\rm out}$ (close), and
ESEL$_{\rm out}$ (wide). 
Notations are same as those in Fig.~\ref{fig:six}, except that there are three lens components 
($M_1$, $M_2$, and $M_3$) and two source stars ($S_1$ and $S_2$).
\smallskip
}
\label{fig:seven}
\end{figure}
% --------------------------------------------------------------

\subsection{ESEL solutions}\label{sec:four-two}

According to the ESEL solutions, the lens and source of the event comprise three lens masses 
($M_1$, $M_2$, and $M_3$) and two source stars ($S_1$ and $S_2$), respectively: 3L2S model.  
Finding solutions is carried out in two steps.  In the first step, we fit the minor anomaly 
by adding an extra lens component: 3L1S model.  Because the overall pattern of the light curve 
is described by a 2L1S model, the anomaly can be treated as a perturbation to the 2L1S model 
\citep{Bozza1999, Han2001}.  Under this assumption, we search for a 3L1S model by fixing the 
parameters of the 2L1S model ($s_2$, $q_2$, and $\alpha$), and then conducting a grid search 
for the parameters describing the third lens component ($s_3$, $q_3$, and $\psi$).  Here 
($s_3$, $q_3$) are the separation and mass ratio of $M_3$ with respect to $M_1$, respectively, 
and $\psi$ indicates the $M_3$  orientation angle, which is measured from the $M_1$--$M_2$ axis 
in a clockwise sense centered at the position of $M_1$.  The parameter ranges are 
$-1.0 \leq \log s_3 \leq 1.0$, $-5.0 \leq \log q_3 \leq 1.0$, and $0 \leq \psi < 2\pi$, and 
they are divided into 50, 50, and 180 grids, respectively.  The solutions are refined by 
gradually narrowing down the ranges of the grid parameters, and then by releasing all parameters 
as free parameters.  In the second step, we fit the major anomaly by introducing an extra source 
$S_2$ based on the 3L1S solution found in the first step. Adding an additional source component 
requires us to include additional parameters, including $(t_{0,2}, u_{0,2}, \rho_2, q_F)$,  
where $q_F$ represents the $S_1/S_2$ flux ratio.

Under the ESEL interpretation, we find four solutions, in which there exist a pair of solutions 
for each of the close and wide solutions.  For all of these solutions, the major anomaly is 
generated by the crossing of $S_2$ over the $M_2$-induced caustic, and the minor anomaly 
is generated by the passage of $S_1$ through the region around the tiny $M_3$-induced caustic.  
Under this interpretation, there are two solutions, designated as ESEL$_{\rm in}$ and ESEL$_{\rm out}$ 
solutions,  for each of the close and wide solutions.  The difference between the ESEL$_{\rm in}$ and 
ESEL$_{\rm out}$ solutions is that the first source ($S_1$) passes the inner side (with respect to 
$M_1$) of the caustic induced by $M_3$ for the ESEL$_{\rm in}$ solution, while $S_1$ passes the outer 
side of the caustic for the ESEL$_{\rm out}$ solution.  Figure~\ref{fig:seven} shows the lensing 
configurations of the four ESEL solutions.

The model curves of the wide ESEL$_{\rm in}$ and ESEL$_{\rm out}$ solutions around the anomalies 
are shown in Figure~\ref{fig:five}, and the parameters of the four ESEL solutions are listed in 
Table~\ref{table:four}.  We note that the mass ratio between $M_3$ and $M_1$ of the lens, 
$q_3\sim (4$--$26)\times 10^{-5}$, is very small, indicating that $M_3$ is a very low-mass planet 
according to these solutions.  Due to the small mass ratio, the caustic induced by $M_3$ is much 
smaller than the caustic induced by $M_2$.

% Table 5 ------------------------------------------------
\begin{deluxetable}{lccc}
\tablecaption{Lensing parameters of ELES models\label{table:five}}
\tablewidth{240pt}
%\tabletypesize{\small}
\tablehead{
\multicolumn{1}{c}{Parameter}           &
\multicolumn{1}{c}{Close}               &
\multicolumn{1}{c}{Wide}                
}
\startdata                           
$\chi^2$                        &   7689.8                    &    7713.7                    \\ 
$t_{0,1}$ (${\rm HJD}^\prime$)  &   $8696.722 \pm 0.012  $    &    $8696.819 \pm 0.009  $    \\
$u_{0,1}$                       &   $0.056 \pm 0.001     $    &    $0.053 \pm 0.001     $    \\ 
$t_{0,2}$ (${\rm HJD}^\prime$)  &   $8695.773 \pm 0.027  $    &    $8696.341 \pm 0.054  $    \\
$u_{0,2}$                       &   $0.147 \pm 0.001     $    &    $0.181 \pm 0.002     $    \\
$\te$ (days)                    &   $43.96 \pm 0.06      $    &    $51.93 \pm 0.08      $    \\
$s_2$                           &   $0.551 \pm 0.003     $    &    $2.524 \pm 0.001     $    \\ 
$q_2$                           &   $0.246 \pm 0.006     $    &    $0.461 \pm 0.002     $    \\
$\alpha$ (rad)                  &   $4.894 \pm 0.005     $    &    $4.897 \pm 0.005     $    \\
$s_3$                           &   $1.054 \pm 0.001     $    &    $0.904 \pm 0.001     $    \\ 
$q_3$ ($10^{-3}$)               &   $4.01 \pm 0.16       $    &    $2.30 \pm 0.10       $    \\
$\psi$ (rad)                    &   $1.348 \pm 0.008     $    &    $1.341 \pm 0.006     $    \\
$\rho_1$ ($10^{-3}$)            &   $0.44 \pm 0.01       $    &    $0.41 \pm 0.01       $    \\
$\rho_2$ ($10^{-3}$)            &    --                       &     --                       \\ 
$q_F$                           &   $0.043 \pm 0.005     $    &    $0.042 \pm 0.005     $    
\enddata                            
%\tablecomments{
%$N_{\rm data}$: number of data points in the data set.
%\smallskip
%}
\end{deluxetable}
% --------------------------------------------------------

% Figure 8 ------------------------------------------------------
\begin{figure}
\includegraphics[width=\columnwidth]{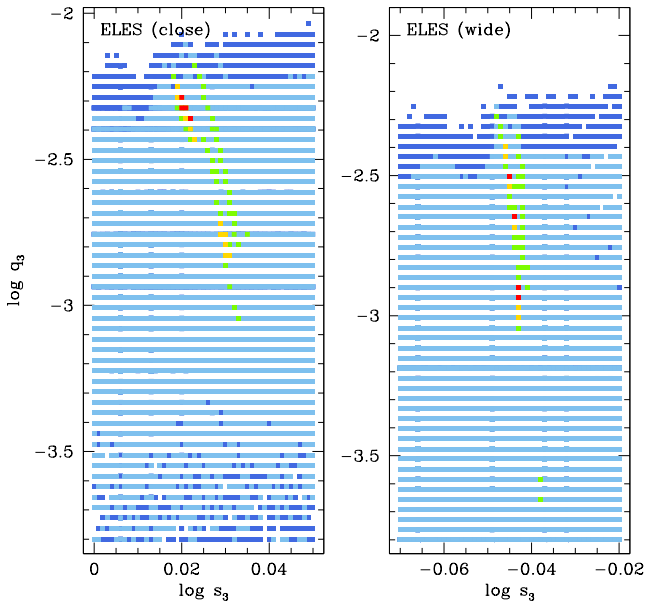}
\caption{
$\Delta\chi^2$ map on the $s_3$--$q_3$ plane for the close and wide 3L1S models. 
The colors of points follows the same coding as in Fig.~\ref{fig:two}.
\smallskip
}
\label{fig:eight}
\end{figure}
% --------------------------------------------------------------

\subsection{ELES solutions}\label{sec:four-three}

We find another 3L2S solutions, in which the way of interpreting the observed anomalies is 
different from that of the 3L2S solutions described in the previous subsection.  We find these 
solutions by first fitting the major anomaly from a 3L1S modeling, and then fitting the minor 
anomaly with the introduction of an extras source $S_2$.  We note that the ESEL solutions, 
discussed in the previous subsection, are obtained by fitting the minor anomaly first from a 
3L1S modeling.  According to the new 3L2S solutions, the major and minor anomalies are explained 
with the addition of $M_3$ and $S_2$ to the baseline 2L1S model, i.e., ELES models.  For each 
of the close and wide solutions, there is no additional degeneracy, and thus there are two ELES 
solutions.  In Figure~\ref{fig:eight}, we present the $\Delta\chi^2$ map on the $s_3$--$q_3$ 
plane.

We list the lensing parameters of the two ELES solutions (close and wide) in Table~\ref{table:five}.  
The estimated mass ratio, $q_3\equiv M_3/M_1\sim (2$--$4)\times 10^{-3}$, is very low, 
suggesting that $M_3$ is a planetary-mass object belonging to the $M_1$--$M_2$ binary system according 
to the ELES model.  The separation and the orientation angle of the planet from $M_1$ is $s_3\sim 1$ 
and $\psi=1.34$ (77.3$^\circ$), respectively.  The value of $\rho_2$ is not presented in the table 
because $S_2$ does not involve with caustic crossings, resulting in no finite effect.

% Figure 9 ------------------------------------------------------
\begin{figure}
\includegraphics[width=\columnwidth]{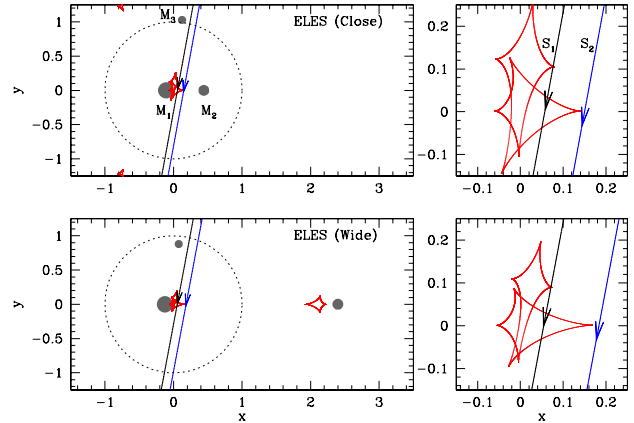}
\caption{
Lensing configurations of the two ELES solutions: ELES (close) and ELES (wide). 
\smallskip
}
\label{fig:nine}
\end{figure}
% --------------------------------------------------------------

The model curve and the residuals from the ELES solutions at around the times of the anomalies 
are shown in Figure~\ref{fig:five}. The lensing configurations are presented in Figure~\ref{fig:nine}.  
It is found that the configuration is greatly different from that of ESEL solutions, although the 
lens systems of both ESEL and ELES solutions constitute the same numbers of the lens and source 
components, i.e., 3L2S model.  We note that the caustics exhibit a self-intersecting pattern, 
which is a characteristic pattern of a 3L system \citep{Gaudi1998}.  From the comparison of the 
lensing configurations with that of the 2L1S model (shown in Figure~\ref{fig:three}), one finds
that $M_3$ induces an additional set of caustics, that overlaps with the one induced by $M_2$.  
According to the solutions, the major anomaly arises due to the crossing of $S_1$ over one tip 
of the caustic induced by $M_3$, while the minor anomaly is produced by the passage of $S_2$ just 
outside of the tip of caustic induced by $M_2$.  The $I$-band flux ratio between $S_1$ and $S_2$ 
is $q_F\equiv F_{S,2}/F_{S,1}\sim 0.042$--0.043, indicating that the second source is a faint star.

\section{Origins of degeneracies}\label{sec:five}

\begin{figure}
\includegraphics[width=\columnwidth]{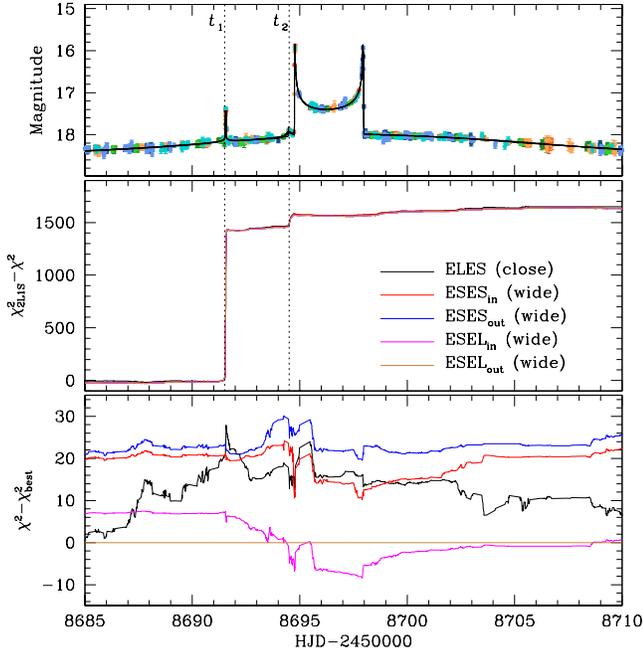}
\caption{
Middle panel:
cumulative distributions of $\Delta\chi^2 = \chi^2_{\rm 2L1S} - \chi^2$ for the five degenerate 
models from the base 2L1S model.  The light curve in the top panel is inserted to  show where the 
fit improves.  The two dotted lines indicate the times of the major ($t_1$) and minor ($t_2$) 
anomalies.  Bottom panel: cumulative distributions of the $\chi^2$ differences from the best-fit 
solution, i.e., $\Delta\chi^2=\chi^2-\chi^2_{\rm ESEL_{out}}$.
}
\label{fig:ten}
\end{figure}

Despite the great differences in the interpretations, it is found that all the degenerate solutions 
provide reasonably good fits to the data.  In Table~\ref{table:six}, we compared the $\chi^2$ values 
of the ten models.  It shows that the close model yields a better fit than the wide model for the 
ELES solutions, while the fits of wide models are better than the fits of the corresponding close 
models for the other solutions.  The middle panel of Figure~\ref{fig:ten} displays the cumulative 
$\Delta\chi^2$ distributions from the base 2L1S model, $\Delta\chi^2=\chi^2-\chi^2_{\rm 2L1S}$, 
for the five solutions, in which we show the distribution of the model providing a better fit 
among each pair of the close and wide solutions.  The distributions show that all models well describe 
both the major and minor anomalies, for which the times of the anomalies, that is, $t_1$ and $t_2$, 
are marked by dotted lines.  The presented distributions of the solutions are very alike, making it 
difficult to distinguish the distributions.  For the better presentation of the differences in the 
fits among the models, we additionally show the distributions of the $\chi^2$ differences from the 
best-fit solution, i.e., $\Delta\chi^2=\chi^2-\chi^2_{\rm ESEL_{out}}$, in the bottom panel.

% Table 6 ------------------------------------------------
\begin{deluxetable}{llll}
\tablecaption{$\chi^2$ values of models\label{table:six}}
\tablewidth{240pt}
%\tabletypesize{\small}
\tablehead{
\multicolumn{1}{c}{Solution}                    &
\multicolumn{2}{c}{$\chi^2$ ($\Delta\chi^2$)}   \\
\multicolumn{1}{c}{}                            &
\multicolumn{1}{c}{Close}                       &
\multicolumn{1}{c}{Wide}          
}
\startdata  
ESES$_{\rm in}$    &  7737.0  (69.1)  &   7696.1  (28.2) \\ 
ESES$_{\rm out}$   &  7771.2  (103.3) &   7701.1  (33.2) \\
ESEL$_{\rm in}$    &  7713.6  (45.7)  &   7668.5  (0.6)  \\
ESEL$_{\rm out}$   &  7741.3  (73.4)  &   7667.9         \\
ELES               &  7689.8  (21.9)  &   7713.7  (45.8) 
\enddata                            
\tablecomments{
The numbers in the parenthesis represent the $\chi^2$ difference from the best-fit solution, 
i.e.,  $\Delta\chi^2=\chi^2-\chi^2_{\rm ESEL_{\rm out}}$, where the best fit is given by the 
wide ESEL$_{\rm out}$ model.
}
\end{deluxetable}
% --------------------------------------------------------

With the multiple interpretations of the event, we probe the origins of the degeneracy. As mentioned, 
the first type arises due to the ambiguity in the separation between $M_1$ and $M_2$, 
i.e., the degeneracy between the close and wide solutions \citep{Griest1998, Dominik1999, Albrow2001}.  
The second type degeneracy arises because the individual short-term anomalies can be described either 
by an extra low-mass lens component or an extra faint source component, i.e, the degeneracy between 
EL and ES solutions.  This degeneracy is similar to the ``planet/binary-source'' degeneracy, that is 
often confronted in interpreting a short-term anomaly superposed on a single-mass light curve, e.g., 
MOA-2012-BLG-486 \citep{Hwang2013}, in the sense that explaining the observed anomaly requires one to 
include an additional lens or source component.  The third degeneracy type is caused by the ambiguity 
between the cusp-approach solution and cusp-crossing solution, i.e., the degeneracy between 
ESES$_{\rm in}$ and ESES$_{\rm out}$ solutions.  This degeneracy arises because the exact trajectory 
of the tertiary source ($S_3$) cannot be specified because the region of the minor anomaly is not densely 
covered.  A similar degeneracy was reported in the interpretation of the short anomaly appeared in 
the OGLE-2015-BLG-1459 light curve \citep{Hwang2018}.  Finally, the ambiguity of $S_1$ trajectory 
relative to the planetary caustic causes an additional degeneracy in the interpretation of the minor 
anomaly.  i.e,. the degeneracy between the ESEL$_{\rm in}$ and ESEL$_{\rm out}$ solutions. This 
degeneracy was mentioned by \citet{Gaudi1997} for a single-host planetary event.  The inspection of 
the degeneracy types reveals that the origins of all the degeneracy types arising in the interpretation 
of the event were already known before.  This implies that checking known types of degeneracies in 
analyzing lensing light curves, especially with complex features, is important to correctly characterize 
the lens system.

% Figure 10 ------------------------------------------------------
% --------------------------------------------------------------

\section{Einstein radius}\label{sec:six}

Measurement of $\thetae$ requires one to estimate the source radius, i.e., $\thetae=\theta_*/\rho$.  
We derive $\theta_*$ from the source color and brightness measured by regressing the 
observed $V$ and $I$ photometric data with the variation of the event magnification.
We then estimate de-reddened values, $(V-I, I)_0$, by calibrating the instrumental 
values, $(V-I, I)$.  This calibration is done utilizing the \citet{Yoo2004} method, in which the 
reference position in the color-magnitude diagram (CMD) for the calibration is the centroid of 
red giant clump (RGC) with $(V-I, I)_{{\rm RGC},0}=(1.060, 14.347)$ known by \citet{Bensby2013} 
and \citet{Nataf2013}.

\begin{figure}[t]
\includegraphics[width=\columnwidth]{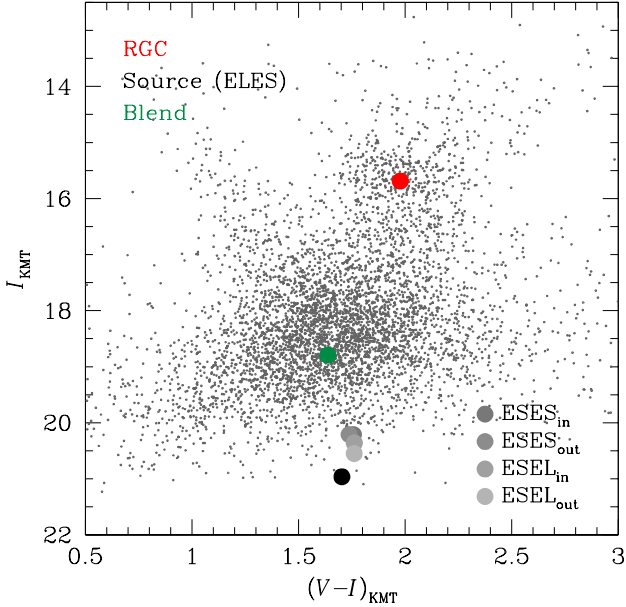}
\caption{
Source, RGC centroid, and blend positions in the CMD.  
The source position is determined based on the ELES solution and those based on the 
other solutions are marked by grey-tone points.
\smallskip
}
\label{fig:eleven}
\end{figure}

In Figure~\ref{fig:eleven}, we mark the source (black dot) in the CMD.  We also mark the RGC 
centroid and blend.  We will discuss in more detail about the nature of the blend in 
Section~\ref{sec:seven}.  We note that the source position is determined using the ELES solution 
for the reason to be discussed in Sect~\ref{sec:seven}.  The source positions based on the other 
solutions are marked by grey-tone points.  The measured color and magnitude are 
$(V-I, I)=(1.703\pm 0.024, 20.963\pm 0.003)$ for the source and 
$(V-I, I)_{\rm RGC} =(1.978, 15.688)$ for the RGC centroid.  With the measured offsets in 
the source color, $\Delta (V-I)$, and magnitude, $\Delta I$, from the RGC centroid, the 
color and magnitude are calibrated as 
\begin{equation}
\eqalign{
(V-I, I)_0 & =  (V-I, I)_{{\rm RGC},0 }+ \Delta (V-I, I) \cr
           & =  (0.786\pm 0.024, 19.622\pm 0.003).       \cr
}
\label{eq1}
\end{equation}
The measured values points out that the source is a main sequence of a late G type.  
We note that the source is well below the brightness limit of {\it Gaia} observation.

In order to estimate the source radius, we convert $V-I$ into $V-K$, and interpolate $\theta_*$ 
from the $(V-K)$--$\theta_*$ relation.  Here we use the \citet{Bessell1988} relation for the 
color conversion, and the \citet{Kervella2004} relation to derive $\theta_*$.  The source is 
estimated to have a radius of 
\begin{equation}
\theta_* = 0.41 \pm 0.03~\mu{\rm as}. 
\label{eq2}
\end{equation}
For $\theta_*$ estimation, the source is assumed to lie at $D_{\rm S}= d_{\rm GC}/[\cos l +\sin l 
(\cos\phi /\sin\phi )]\sim 7.85$~kpc, where we adopt a Galactocentric distance of $d_{\rm GC}=8.16$~kpc 
and a bulge bar orientation angle of $\phi= 40^\circ$.  The error of the $\theta_*$ measurement is 
estimated based on the error of the measured source color and adding 
7\% error in quadrature to account for combined uncertainty resulting from the RGC centroiding and 
the color-$\theta_*$ conversion.  We note that the source companion contributes little flux, 
$q_F/(1+q_F)\sim 4\%$, to the combined source flux, and thus its effect on the estimated $\theta_*$ 
is minimal.  Together with $\rho$ and $\te$, the measured $\theta_*$ yields the Einstein radius and 
the relative lens-source proper motion of
\begin{equation}
\thetae = {\theta* \over \rho} = 0.91 \pm 0.07~{\rm mas}, 
\label{eq3}
\end{equation}
and
\begin{equation}
\mu = {\thetae \over \te} = 7.59 \pm 0.59~{\rm mas}~{\rm yr}^{-1}, 
\label{eq4}
\end{equation}
respectively. The estimated value of $\thetae$ is substantially bigger than typical 
value of $\sim 0.5$~mas for events with low-mass stellar lenses lying about halfway between Earth 
and the bulge. This suggests that the lens 
lies at a close distance.

% Figure 11 ------------------------------------------------------
% --------------------------------------------------------------

\section{Resolving Degeneracy}\label{sec:seven}

We check whether the degeneracies among the solutions can be resolved.  For this, we compare the
flux ratio between $S_1$ and $S_2$ measured from modeling, $q_{F,{\rm model}}=F_{S,2}/F_{S,1}$, with 
the flux ratio predicted from the radius ratio $\rho_2/\rho_1$, $q_{F,{\rm pred}}$. The radius ratio 
$\rho_2/\rho_1$ is measurable for the solutions in which the major anomaly is explained by the caustic 
crossing of $S_2$, i.e., ESXX models. We note that the major anomaly for the ELES model is explained by 
the caustic crossing of $S_1$ instead of $S_2$, and thus this method cannot be applied. In order to 
estimate $q_{F,{\rm pred}}$ from the $\rho_2/\rho_1$ ratio, we first estimate the physical radius of 
$S_2$ by
\begin{equation}
R_{*,S_2} =\left( {\rho_2 \over \rho_1}\right) R_{*,S_1}.
\label{eq5}
\end{equation}
The ESXX solutions result in similar values of the color and brightness, as shown in Figure~\ref{fig:eleven}, 
and thus we use a common physical source radius of $R_{*,S_1}\sim 0.85~R_\odot$, which is deduced from 
$(V-I)_0$ and $I_0$.  We then estimate the absolute $I$-band magnitudes of the source 
stars, $M_{I,1}$ and $M_{I,2}$, corresponding to $R_{*,S_1}$ and $R_{*,S_2}$ from the tables of 
\citet{Pecaut2012} and \citet{Pecaut2013}, and compute the flux ratio between $S_1$ and $S_2$ as 
\begin{equation}
q_{F,{\rm pred}} = 10^{-0.4(M_{I,2}-M_{I,1})}.
\label{eq6}
\end{equation}
We note that a similar method could, in principle, be applied to the flux ratio $q_{F,3}=F_{S,3}/F_{S,1}$ for 
the models with three source stars, i.e., ESES solutions. However, the normalized radius of the tertiary 
source, $\rho_3$, is measured for none of the ESES model, and thus the method cannot be implemented.

% Figure 12 ------------------------------------------------------
\begin{figure}
\includegraphics[width=\columnwidth]{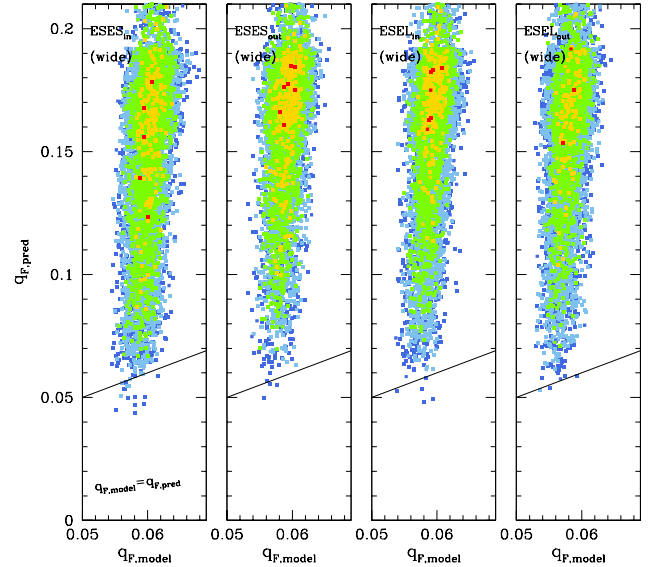}
\caption{
Plot of $q_{F,{\rm pred}}$ versus $q_{F,{\rm model}}$ for the models explaining the major anomaly 
by introducing an extra source.  Color codings represent points with $\leq 1\sigma$ (red), 
$\leq 2\sigma$ (yellow), $\leq 3\sigma$ (green), $\leq 4\sigma$ (cyan), and $\leq 5\sigma$ (blue).  
The oblique solid line in each panel represents the positions at which $q_{F,{\rm pred}}=
q_{F,{\rm model}}$.
}
\label{fig:twelve}
\end{figure}
% --------------------------------------------------------------

Figure~\ref{fig:twelve} shows the scatter plot of points in the MCMC chains on the 
$q_{F,{\rm model}}$--$q_{F,{\rm pred}}$ parameter plane for the four tested ESXX models.  In 
the plot, the color coding represents points with $\leq 1\sigma$ (red), $\leq 2\sigma$ (yellow), 
$\leq 3\sigma$ (green), $\leq 4\sigma$ (cyan), and $\leq 5\sigma$ (blue).  The oblique solid line 
in each panel represents the positions at which $q_{F,{\rm pred}}=q_{F,{\rm model}}$.  The plots 
show that the hypothesis of the second source's caustic crossing for the origin of the major 
anomaly is rejected at more than $3\sigma$ level, suggesting that the ESXX models are unlikely 
to be correct interpretations of the event.  With these solutions rejected, the ELES solutions 
remain as the only viable interpretation of the event.

\section{Physical lens parameters}\label{sec:eight}

Recognizing that only the ELES models provide plausible interpretations of the observed lensing
data, the lens parameters of the lens mass, $M$, and distance, $\dl$, are estimated 
based on the observables of the ELES solutions.  In order to unambiguously determine these parameters, 
one should measure both $\thetae$ and $\pie$, where $\pie$ represents the microlens parallax 
(hereafter parallax), i.e., 
\begin{equation}
M={\thetae \over \kappa\pie};\qquad \dl = {{\rm au} \over \pie\thetae + \pi_{\rm S}}.
\label{eq7}
\end{equation}
Here $\kappa =4G/(c^2{\rm au})$, $D_{\rm S}$ is the source distance, and $\pi_{\rm S}={\rm au}/D_{\rm S}$.  
The Einstein radius is firmly measured, i.e., Equation~(\ref{eq3}).  An additional modeling conducted 
considering the parallax effect indicates that secure determinations of the parallax parameters are 
difficult.  Although $\pie$ is not measured, the lensing observables of $\te$ and $\thetae$ are 
related as 
\begin{equation}
\te = {\thetae \over \mu};\qquad
\thetae = (\kappa M \pi_{\rm rel})^{1/2},
\label{eq8}
\end{equation}
where $\pi_{\rm rel}={\rm au}(D_{\rm L}^{-1} - D_{\rm S}^{-1})$.  Then, the lens parameters 
can be estimated by conducting a Bayesian analysis  with the prior Galactic model, defining the 
mass function, physical, and dynamical distributions.

% Figure 13 ------------------------------------------------------
\begin{figure}
\includegraphics[width=\columnwidth]{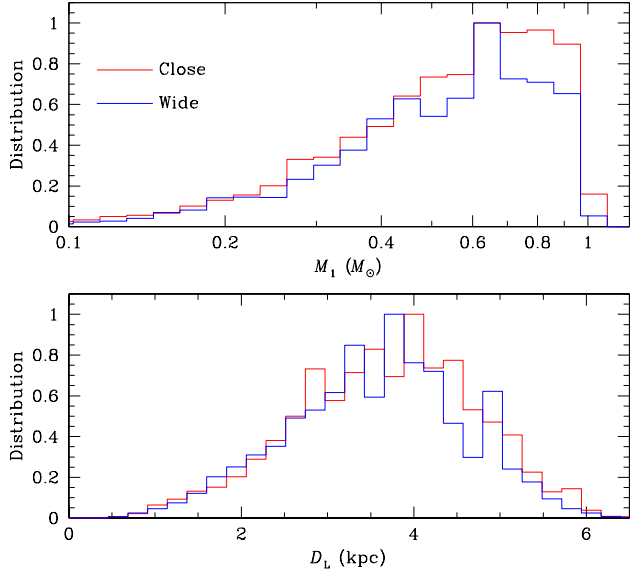}
\caption{
Bayesian posterior distributions of $M_1$ and $\dl$.  The red and blue curves represent the 
distributions corresponding to the close and wide models, respectively.
}
\label{fig:thirteen}
\end{figure}
% --------------------------------------------------------------

The Bayesian analysis is carried out in two steps.  The first step is producing events from 
a Monte Carlo simulation using a Galactic model.
The models used for the analysis include \citet{Han2003}, \citet{Han1995}, \citet{Zhang2020} 
models for the physical, dynamical distributions, and mass function, respectively.  See section~5 
of \citet{Han2020b} for more details about the models.  In the second step, the posterior 
distributions of $M$ and $\dl$ are constructed for the events with observables, i.e., $\te$ 
and $\thetae$, in the 1$\sigma$ ranges of the estimated observables among the events produced 
from the simulation.

The posterior distributions $M_1$ and $\dl$ derived from the ELES solutions are shown in 
Figure~\ref{fig:thirteen}, where the red curve is the distribution resulting from the close 
solution, while the blue curve is the one from the wide solution.  In Table~\ref{table:seven}, 
we summarize the estimated parameters of $M_1$, $M_2$, $M_3$, $\dl$, $d_{\perp,2}$, and 
$d_{\perp,3}$, where the last two quantities are the separations of $M_2$ and $M_3$ from $M_1$.  
The listed values are the medians of the posterior distributions with errors determined as 16\% 
and 84\% ranges.  According to the estimated parameters from the close model, the lens is a 
planetary system, in which a giant planet with a mass 
\begin{equation}
M_p = 2.56^{+1.13}_{-1.16}~M_{\rm J}
\label{eq9}
\end{equation}
belongs to a stellar binary composed of K and M dwarfs with masses 
\begin{equation}
M_1 =0.61^{+0.27}_{-0.28}~M_\odot,
\label{eq10}
\end{equation}
and
\begin{equation}
M_2 = 0.15^{+0.07}_{-0.07}~M_\odot,
\label{eq11}
\end{equation}
respectively.  As predicted by the large $\thetae$, the lens is located at a close distance of
\begin{equation}
\dl = 3.86^{+1.00}_{-1.12}~{\rm kpc},
\label{eq12}
\end{equation}
suggesting that it is a disk object. The lens parameters derived from the wide 
model are also listed in Table~\ref{table:seven}.  According to the ELES models, the lens 
is the seventh microlensing system with a planet in a stellar binary, followed by 
OGLE-2006-BLG-284L \citep{Bennett2020}, 
OGLE-2007-BLG-349L \citep{Bennett2016},
OGLE-2008-BLG-092L \citep{Poleski2014}, 
OGLE-2013-BLG-0341L \citep{Gould2014}, 
OGLE-2016-BLG-0613L \citep{Han2017}, and 
OGLE-2018-BLG-1700L \citep{Han2020a}.

% Table 7 ------------------------------------------------
\begin{deluxetable}{lccc}
\tablecaption{Physical lens parameters\label{table:seven}}
\tablewidth{240pt}
%\tabletypesize{\small}
\tablehead{
\multicolumn{1}{c}{Parameter}         &
\multicolumn{1}{c}{Close}             &
\multicolumn{1}{c}{Wide}              
}
\startdata  
$M_1$ ($M_\odot$)     &   $0.61^{+0.27}_{-0.28}$   &    $0.59^{+0.26}_{-0.26}$  \\
$M_2$ ($M_\odot$)     &   $0.15^{+0.07}_{-0.07}$   &    $0.27^{+0.12}_{-0.12}$  \\ 
$M_3$ ($M_{\rm J}$)   &   $2.56^{+1.13}_{-1.16}$   &    $1.43^{+0.62}_{-0.62}$  \\
$\dl$ (kpc)           &   $3.86^{+1.00}_{-1.12}$   &    $3.77^{+0.98}_{-1.13}$  \\
$d_{\perp,2}$ (au)    &   $1.73^{+0.43}_{-0.48}$   &    $7.82^{+2.04}_{-2.35}$  \\
$d_{\perp,3}$ (au)    &   $3.32^{+0.82}_{-0.92}$   &    $2.80^{+0.73}_{-0.84}$  
\enddata                            
%\tablecomments{
%$N_{\rm data}$: number of data points in the data set.
%\smallskip
%}
\end{deluxetable}
% --------------------------------------------------------

The estimated physical lens parameters suggest the possibility that the majority of the 
blended light comes from the lens.  In Figure~\ref{fig:eleven}, We mark the blend, with 
$(V-I, I)_b\sim (1.6, 18.8)$, in the CMD.  According to the estimated $M$ and $\dl$, the 
predicted brightness and color of the primary lens, which explains most of the flux from 
the lens, are in the ranges of 
\begin{equation}
18.3 \lesssim I_{\rm L} \lesssim  22.3 
\label{eq13}
\end{equation}
and
\begin{equation}
1.3 \lesssim (V-I)_{\rm L} \lesssim  2.9.
\label{eq14}
\end{equation}
Here we compute the lens brightness and color by $I_{\rm L}=M_I + 5 \log \dl - 5 + A_I$ and 
$(V-I)_{\rm L}=(V-I)_0 + E(V-I)$, where $M_I$ and $(V-I)_0$ are the absolute $I$-band magnitude 
and intrinsic color corresponding to $M_1$, and we assume $A_I\sim 0.5$ and $E(V-I)\sim 0.4$ 
by adopting the half of the values to the source considering $D_{\rm L}$.  Then, the color and 
brightness of the blend are within the predicted ranges of the lens.

We further check this hypothesis by measuring the offset between the centroid of the apparent 
source image at the baseline and the source position in the difference image during the lensing 
magnification.  We measure 
\begin{equation}
(\Delta x, \Delta y)=(0.21\pm 0.10, 0.19\pm 0.18)~{\rm arcsec}. 
\label{eq15}
\end{equation}
Considering that the measured offset is within $\sim 2\sigma$ of the measurement error, the 
hypothesis cannot be ruled out.  Therefore, an important portion of the blended light may 
come from the lens, but this can only be established using high-resolution images that can 
be obtained from future observations with the use of adaptive optics (AO) instrument mounted 
on 8m-class telescopes or space-based telescopes.

\section{Summary}\label{sec:nine}

We present the result from the investigation of  KMT-2019-BLG-1715.  The event light curve 
displayed two anomalies from a typical caustic-crossing binary-lensing light curve.  We 
identified five pairs of solutions, in which the anomalies were explained with the inclusion 
of an extra lens or source component in addition to the base binary-lens model.  We presented 
detailed analysis for the individual solutions, and trace the origins of degeneracies.  To 
resolve the degeneracies, we compare the measured $S_1/S_2$ flux ratio with the ratio deduced 
from the ratio of the source radii.  Applying this method left only a single pair of viable 
solutions, in which the anomaly with a large deviation was produced by a third body of the 
lens, and the anomaly with a small deviation was generated by a second source.  A Bayesian 
analysis indicated that the lens comprised a $\sim 2.6~M_{\rm J}$ planet and binary stars with 
K and M spectral types lying in the disk.   We pointed out that the lens might be the blend, 
and this hypothesis could be confirmed from future high-resolution followup observations.

\acknowledgments
Work by C.H.\ was supported by the grants  of National Research Foundation of Korea 
(2020R1A4A2002885 and 2019R1A2C2085965)
% Gould  
Work by A.G.\ was supported by JPL grant 1500811.
% OGLE  
The OGLE project has received funding from the National Science Centre, Poland, grant
MAESTRO 2014/14/A/ST9/00121 to A.U..
% MOA 
The MOA project was supported by the grant 19KK0082 and 20H04754.
Work by Y.H, was supported by JSPS KAKENHI grant No.~17J02146. 
D.P.B., A.B., and C.R. were supported by NASA through grant NASA-80NSSC18K0274. 
Work by N.K. was supported by JSPS KAKENHI grant No.~JP18J008.
Work by C.R.\ was supported by an appointment to the NASA Postdoctoral Program at the Goddard Space Flight Center,
administered by USRA through a contract with NASA.
% KMTNet
This research has made use of the KMTNet system operated by the Korea
Astronomy and Space Science Institute (KASI) and the data were obtained at
three host sites of CTIO in Chile, SAAO in South Africa, and SSO in Australia.


\begin{thebibliography}{}
% -------------
\bibitem[Alard \& Lupton(1998)]{Alard1998} Alard, C., \& Lupton, R.~H.\ 1998, \apj, 503, 325
\bibitem[Albrow(2017)]{Albrow2017} Albrow, M.\ 2017, MichaelDAlbrow/pyDIA: Initial Release on Github,Version v1.0.0, Zenodo, doi:10.5281/zenodo.268049
\bibitem[Albrow et al.(2001)]{Albrow2001} Albrow, M.~D., An, J., Beaulieu, J.-P., et al.\ 2001, \apj, 549, 759  
\bibitem[Alcock et al(1997)]{Alcock1997} Alcock, C., Allsman, R.~A., Alves, D., et al.\ 1997, \apj, 479, 119
\bibitem[An \& Han(2002)]{An2002}  An, J.~H., \& Han, C.\ 2002, \apj, 573, 351
\bibitem[Bennett(2010)]{Bennett2010} Bennett, D.~P.\ 2010, \apj, 716, 1408
\bibitem[Bennett et al.(2016)]{Bennett2016} Bennett, D.~P., Rhie, S.~H., Udalski, A., et al.\ 2016, \aj, 152, 125
\bibitem[Bennett et al.(2020)]{Bennett2020} Bennett, D.~P., Udalski, A., Bond, I.~A., et al.\ 2020, \aj, 160, 72 
\bibitem[Bensby et al.(2013)]{Bensby2013} Bensby, T., Yee, J.~C., Feltzing, S., et al.\ 2013, \aap, 549, 147
\bibitem[Bessell \& Brett(1988)]{Bessell1988} Bessell, M.~S., \& Brett, J.~M. 1988, \pasp, 100, 1134
\bibitem[Bond et al.(2001)]{Bond2001} Bond, I. A., Abe, F., Dodd, R.~J., et al.\ 2001, \mnras, 327, 868
\bibitem[Bond et al.(2002)]{Bond2002} Bond, I.~A., Rattenbury, N.~J., Skuljan, J., et al.\ 2002, \mnras, 333, 71 
\bibitem[Bozza(1999)]{Bozza1999} Bozza, V.\ 1999, \aap, 348, 311
\bibitem[Bozza et al.(2018)]{Bozza2018} Bozza, V., Bachelet, E., Bartoli\'c, F., Heintz, T.~M., Hoag, A.~R., \& Hundertmark, M.\ 2018, \mnras, 479, 5157
\bibitem[Bozza et al.(2012)]{Bozza2012} Bozza, V., Dominik, M., Rattenbury, N.~J., et al.\ 2012, \mnras, 424, 902
%\bibitem[Gaia Collaboration(2018)]{Gaia2018} Gaia Collaboration, Brown, A.~G.~A., Vallenari, A., et al.\ 2018, \aap, 616, A1
\bibitem[Dan\v{e}k \& Heyrovsk\'y(2015)]{Danek2015} Dan\v{e}k, K., \& Heyrovsk\'y, D.\ 2015, \apj, 806, 99
\bibitem[Dan\v{e}k \& Heyrovsk\'y(2019)]{Danek2019} Dan\v{e}k, K., \& Heyrovsk\'y, D.\ 2019, \apj, 880, 72
\bibitem[Di Stefano \& Mao(1996)]{Stefano1996}  Di Stefano, R., \& Mao, S. 1996, \apj, 457, 93
\bibitem[Dominik(1999)]{Dominik1999} Dominik, M.\ 1999, \aap, 349, 108 
\bibitem[Dong et al.(2009)]{Dong2009} Dong, S., Bond, I.~A., Gould, A., et al. 2009, \apj, 698, 1826
\bibitem[Gaudi(1998)]{Gaudi1998}  Gaudi, B.~S.\ 1998, \apj, 506, 533
\bibitem[Gaudi et al.(1998)]{Gaudi1998} Gaudi, B.~S., Naber, R.~M., \& Sackett, P.~D.\ 1998, \apjl, 502, L33
\bibitem[Gaudi \& Gould(1997)]{Gaudi1997} Gaudi, B.~S., \& Gould A. 1997, \apj, 486, 85
\bibitem[Gaudi \& Han(2004)]{Gaudi2004}  Gaudi, B.~S., \& Han, C.\ 2004, \apj, 611, 528
\bibitem[Gould(1992)]{Gould1992} Gould, A.\ 1992, \apj, 392, 442
\bibitem[Gould \& Gaucherel(1997)]{Gould1997} Gould, A., \& Gaucherel, C.\ 1997, \apj, 477, 580 
\bibitem[Gould et al.(2014)]{Gould2014} Gould, A., Udalski, A., Shin, I.-G., et al.\ 2014, Science, 345, 46
\bibitem[Griest \& Safizadeh(1998)]{Griest1998} Griest, K., \& Safizadeh, N.\ 1998, \apj, 500, 37
\bibitem[Han et al.(2001)]{Han2001}  Han, C., Chang, H.-Y., An, J. H., \& Chang, K.\ 2001, \mnras, 328, 986
\bibitem[Han \& Gould(1995)]{Han1995} Han, C., \& Gould, A.\ 1995, \apj, 447, 53
\bibitem[Han \& Gould(2003)]{Han2003} Han, C., \& Gould, A.\ 2003, \apj, 592, 172
\bibitem[Han et al.(2021)]{Han2021} Han, C., Lee, C.-U., Ryu, Y.-H., et al.\ 2021, \aap, in press [arXiv:2102.01806]
\bibitem[Han et al.(2020a)]{Han2020a} Han, C., Lee, C.-U., Udalski, A., et al.\ 2020a, \aj, 159, 48
\bibitem[Han et al.(2020b)]{Han2020b} Han, C., Shin, I.-G., Jung, Y. K., et al.\ 2020b, \aap, 641, A105 
\bibitem[Han et al.(2017)]{Han2017} Han, C., Udalski, A., Gould, A., et al.\ 2017, \aj, 154, 223
\bibitem[Hwang et al.(2013)]{Hwang2013} Hwang, K. -H., Choi, J. -Y., Bond, I. A., et al.\ 2013\ \apj, 778, 55 
\bibitem[Hwang et al.(2018)]{Hwang2018} Hwang, K.-H., Udalski, A., Bond, I.~A., et al.\ 2018, \aj, 155, 259
\bibitem[Kervella et al.(2004)]{Kervella2004} Kervella, P., Th\'evenin, F., Di Folco, E., \& S\'egransan, D.\ 2004, \aap, 426, 29
\bibitem[Kim et al.(2016)]{Kim2016} Kim, S.-L., Lee, C.-U., Park, B.-G., et al.\ 2016, JKAS, 49, 37
\bibitem[Mao \& Paczy\'nski(1991)]{Mao1991} Mao, S., \& Paczy\'nski, B.\ 1991, \apjl, 374, 37
\bibitem[Nataf et al.(2013)]{Nataf2013} Nataf, D.~M., Gould, A., Fouqu\'e, P., et al.\ 2013, \apj, 769, 88
\bibitem[Pecaut \& Mamajek(2013)]{Pecaut2013} Pecaut, M.~J. \& Mamajek, E. E. 2013, \apjs, 208, 9
\bibitem[Pecaut et al.(2012)]{Pecaut2012} Pecaut, M.~J., Mamajek, E.~E., \& Bubar, E.~J. 2012, \apj, 746, 154
\bibitem[Poleski et al.(2014)]{Poleski2014} Poleski, R., Skowron, J., Udalski, A., et al.\ 2014, \apj, 795, 42
\bibitem[Ryu et al.(2010)]{Ryu2010} Ryu, Y. -H., Han, C., Hwang, K.-H., et al.\ 2010, \apj, 723, 81
%\bibitem[Tomaney \& Crotts(1996)]{Tomaney1996} Tomaney, A.~B., \& Crotts, A.~P.~S.\ 1996, \aj, 112, 2872
\bibitem[Udalski(2003)]{Udalski2003} Udalski,~A.\ 2003, Acta Astron., 53, 291
\bibitem[Udalski et al.(2015)]{Udalski2015} Udalski,~A., Szyma\'nski,~M.~K., \& Szyma\'nski,~G.\ 2015, Acta Astron., 65, 1
\bibitem[Udalski et al.(1994a)]{Udalski1994a} Udalski, A., Szyma\'nski, M., Ka{\l}u\.zny, J., Kubiak, M., Mateo, M., \& Krzemi\'nski, W.\ 1994, Acta Astron. 44, 1
\bibitem[Udalski et al.(1994b)]{Udalski1994b} Udalski, A., Szyma\'nski, M., Mao, S., Di Stefano, R., Ka{\l}u\.zny, J., Kubiak, M., Mateo, M., \& Krzemi\'nski, W. 1994, \apjl, 436, L103
\bibitem[Wo\'zniak(2000)]{Wozniak2000} Wo\'zniak, P. R.\ 2000, Acta Astron., 50, 42
\bibitem[Yee et al.(2012)]{Yee2012} Yee, J.~C., Shvartzvald, Y., Gal-Yam, A., et al.\ 2012, \apj, 755, 102
\bibitem[Yoo et al.(2004)]{Yoo2004} Yoo, J., DePoy, D.~L., Gal-Yam, A., et al.\ 2004, \apj, 603, 139
\bibitem[Zhang et al.(2020)]{Zhang2020} Zhang, X., Zang, W., Udalski, A., et al.\ 2020, \aj, 159, 116
\end{thebibliography}
\end{document}